\documentclass{ws-procs975x65}

\begin{document}

\title{PROPERTIES OF A TWO-SPHERE SINGULARITY}  

\author{DEBORAH A. KONKOWSKI}

\address{Department of Mathematics, U.S. Naval Academy\\
Annapolis, Maryland, 21012, USA\\
E-mail: dak@usna.edu}

\author{THOMAS M. HELLIWELL}

\address{Department of Physics, Harvey Mudd College\\
Claremont, California, 91711, USA\\
E-mail: helliwell@HMC.edu}

\begin{abstract}
Recently B\"ohmer and Lobo have shown that a metric due to
Florides can be extended to reveal a classical singularity that has the form of a
two-sphere. Here we discuss and expand on the classical singularity
properties and then show the classical singularity is not healed by
a quantum analysis.

\end{abstract}

\bodymatter

\section{Introduction}
The question addressed in this review is: What are the properties of the unusual two-sphere singularity discovered by B\"ohmer and Lobo? The answer is given for quantum as well as classical singularity structure. This conference proceeding is based on articles by B\"ohmer and Lobo \cite{BL} and by the authors \cite{HK1, HK2}.

\section{Types of Singularities}

\subsection{Classical Singularities}
A classical singularity is indicated by incomplete geodesics or incomplete paths of bounded acceleration \cite{HE, Geroch} in a maximal spacetime. Since, by definition, a spacetime is smooth, all irregular points (singularities) have been excised; a singular point is a boundary point of the spacetime. There are three different types of singularity \cite{ES}: quasi-regular, non-scalar curvature and scalar curvature. Whereas quasi-regular singularities are topological, curvature singularities are indicated by diverging components of the Riemann tensor when it is evaluated in a parallel-propagated orthonormal frame carried along a causal curve ending at the singularity.

\subsection{Quantum Singularities}
A spacetime is QM (quantum-mechanically) nonsingular if the evolution of a test scalar wave packet, representing the quantum particle, is uniquely determined by the initial wave packet, manifold and metric, without having to put boundary conditions at the singularity\cite{HM}. Technically, a static ST (spacetime) is QM-singular if the spatial portion of the Klein-Gordon operator is not essentially self-adjoint on $C_{0}^{\infty}(\Sigma)$ in $L^2(\Sigma)$ where $\Sigma$ is a spatial slice. This is tested (see, e.g., Konkowski and Helliwell \cite{HK1, HK2, KH}) using Weyl's limit point - limit circle criterion \cite{RS, Weyl} that involves looking at an effective potential asymptotically at the location of the singularity. Here a limit-circle potential is quantum mechanically singular, while a limit-point potential is quantum mechanically non-singular.

\section{2-Sphere Singularity -- B\"ohmer-Lobo Space-time}

\par The B\"ohmer and Lobo metric  \cite{BL} is

\begin{equation}
ds^2 = -\frac{dt^2}{\cos\alpha} + R^2 d\alpha^2 + R^2 \sin^2\alpha \  d\Omega^2.
\end{equation}

\noindent where $R = \sqrt{3/8 \pi \rho_0}$ in terms of the constant energy density $\rho_0$, and $d\Omega^2 = d\theta^2 + \sin^2 \theta d\phi^2$. The coordinate ranges are $- \infty < t < \infty $, $0 \le \theta \le \pi$,  and $0 \le \phi < 2\pi$. The radial coordinate $\alpha$ can either take the values $0 < \alpha \le \pi/2$ (half a three-sphere) or  $- \pi/2 \le \alpha \le \pi/2$ (two half three-spheres joined at $\alpha = 0$ with $\alpha = - \pi /2$ identified with $\alpha = + \pi/2.$).

\par The B\"ohmer-Lobo spacetime is static, spherically symmetric, regular at $\alpha = 0$, and it has vanishing radial stresses \cite{BL}. It is also Petrov Type D and Segre Type A1 ([(11) 1, 1]), and it satisfies the strong energy condition automatically and the dominate energy condition with certain more stringent requirements \cite{BL}.  Vertical cuts through the three-sphere define latitudinal two-spheres; in particular, the equatorial cut at $\alpha = \pi/2$ is a two-sphere on which scalar polynomial invariants diverge and the tangential pressure diverges as well.

\subsection{Classical singularity structure}

\par One can show that the B\"ohmer-Lobo spacetime is timelike geodesically complete but null geodesically incomplete. The equatorial two-sphere is a weak, timelike, scalar curvature singularity.

\subsection{Quantum singularity structure}

The Klein-Gordon equation

\begin{equation}
|g|^{-1/2}\left(|g|^{1/2}g^{\mu \nu} \Phi,_{\nu}\right),_{\mu} = M^2 \Phi
\end{equation}

\noindent for a scalar function $\Phi$ has mode solutions of the form

\begin{equation}
\Phi \sim e^{- i \omega t} F(\alpha) Y_{\ell m}(\theta, \phi)
\end{equation}

\noindent for spherically symmetric metrics, where the $Y_{\ell m}$ are spherical harmonics and $\alpha$ is the radial coordinate. The radial function $F(\alpha)$ for the B\"ohmer-Lobo metric obeys

\begin{equation}
F'' + \left(2\cot\alpha + \frac{1}{2}\tan\alpha\right) F' + \left[R^2 \omega^2 \cos\alpha - \frac{\ell(\ell + 1)}{\sin^2\alpha} - R^2 M^2\right] F = 0,
\end{equation}

\noindent and square integrability is judged by finiteness of the integral

\begin{equation}
I = \int d\alpha d\theta d\phi \sqrt{\frac{g_3}{g_{00}}} \Phi^* \Phi ,
\end{equation}

\noindent where $g_3$ is the determinant of the spatial metric. A change of coordinates puts the singularity at $x=0$ and converts the integral and differential equation to the one-dimensional Schr\"odinger forms $\int dx \psi^* \psi$ and

\begin{equation}
\frac{d^2 \psi}{dx^2} + (E - V)\psi = 0,
\end{equation}

\noindent where $E = R^2 \omega^2$ with a potential that is asymptotically

\begin{equation}
V(x) \sim \frac{R^2 M^2 + \ell (\ell + 1)}{x^{2/3}} < \frac{3}{4x^2}.
\end{equation}

\noindent It follows from Theorem X.10 in Reed and Simon \cite{RS} that $V(x)$ is in the limit circle case, so $x = 0$ is a quantum singularity. The Klein-Gordon operator is therefore not essentially self-adjoint. Quantum mechanics fails to heal the singularity.

\bibliographystyle{ws-procs975x65}
\bibliography{ws-pro-sample}

\begin{thebibliography}{9}
\bibitem{BL} C.G. B\"ohmer and F.S.N. Lobo, {\em Int. J. Mod. Phys. D}{\bf 17}, 897 (2008).
\bibitem{HK1} T.M. Helliwell and D.A. Konkowski, {\em Gen. Rel. Grav.} {\bf 43}, 695 (2011).
\bibitem{HK2} T.M. Helliwell and D.A. Konkowski, {\em Int. J. Mod. Phys. A} {\bf 26}, 3878 (2011).
\bibitem{HE} S.W. Hawking and G.F.R. Ellis, {\em The Large-Scale Structure of Spacetime} (Cambridge University Press, 1973).
\bibitem{Geroch}R. Geroch, {\em Ann. Phys.} {\bf 48}, 526 (1968) 
\bibitem{ES} G.F.R. Ellis and B.G. Schmidt, {\em Gen. Rel. Grav.} {\bf 8}, 915 (1977).
\bibitem{HM} G.T. Horowitz and D. Marolf, {\em Phys. Rev. D} {\bf 52}, 5670 (1995).
\bibitem{KH} D.A. Konkowski and T.M. Helliwell, {\em Gen. Rel. Grav.} {\bf 38}, 1069 (2006).
\bibitem{RS} M. Reed and B. Simon, {\em Functional Analysis} (Academic Press, 1972); M. Reed and B. Simon, {\em Fourier Analysis and Self-Adjointness} (Academic Press, 1972). 
\bibitem{Weyl} H. Weyl, {\em Math. Ann.} {\bf 68}, 220 (1910).
\end{thebibliography}

\end{document}